\documentclass[longauth,traditabstract,letter]{aa}
\pdfoutput=1

\usepackage{graphicx}
\usepackage{txfonts}
\usepackage{natbib}
\bibpunct{(}{)}{;}{a}{}{,}
\usepackage{hyperref}
\hypersetup{pdfauthor={M. de Val-Borro et al.},pdftitle={Water production in comet 81P/Wild 2 as determined by Herschel/HIFI}}
\usepackage{ucs}
\usepackage[utf8x]{inputenc}
\usepackage[T1]{fontenc}
\usepackage{tikzexternal}
\newcommand{\pgfplotsset}[1] {}
\newcommand{\wild}{81P/Wild~2}
\newcommand{\herschel}{{\it Herschel}}
\newcommand{\stardust}{{\it Stardust}}
\newcommand{\xne}{x_{n_\mathrm{e}}}
\newcommand{\rh}{r_\mathrm{h}}
\newcommand{\vexp}{v_\mathrm{exp}}
\newcommand{\range}{0.9\textrm{--}1.1}
\newcommand{\mean}{1.0}
\newcommand{\transi}{$1_{10}$--$1_{01}$}
\newcommand{\transii}{$1_{11}$--$0_{00}$}
\newcommand{\ortho}{556.936}
\newcommand{\para}{1113.343}
\newcommand{\orthos}{557}
\newcommand{\paras}{1113}
\newcommand{\tlabel}{$T_\mathrm{mB}$ [K]}
\newcommand{\vlabel}{$v$ [\kms]}
\newcommand\water{\ifmmode{\rm H_2O}\else H$_2$O\fi}
\newcommand\oh{\ifmmode{\rm OH}\else OH\fi}
\newcommand\kms{\ifmmode{\rm km\thinspace s^{-1}}\else km\thinspace s$^{-1}$\fi}
\newcommand\ms{\ifmmode{\rm m\thinspace s^{-1}}\else m\thinspace s$^{-1}$\fi}
\newcommand\s{\ifmmode{\rm s^{-1}}\else s$^{-1}$\fi}
\graphicspath{{./figures/}}
\newenvironment{DIFnomarkup}{}{}

\begin{document}

\title{Water production in comet \wild{} as determined by \herschel{}/HIFI%
\thanks{\herschel{} is an ESA space observatory with science instruments
provided by European-led Principal Investigator consortia and with
important participation from NASA.}}

\author{
  M.~de~Val-Borro\inst{\ref{inst1}}
  \and P.~Hartogh\inst{\ref{inst1}}
  \and J.~Crovisier\inst{\ref{inst2}}
  \and D.~Bockel\'ee-Morvan\inst{\ref{inst2}}
  \and N.~Biver\inst{\ref{inst2}}
  \and D.~C.~Lis\inst{\ref{inst3}}
  \and R.~Moreno\inst{\ref{inst2}}
  \and C.~Jarchow\inst{\ref{inst1}}
  \and M.~Rengel\inst{\ref{inst1}}
  \and S.~Szutowicz\inst{\ref{inst4}}
  \and M.~Banaszkiewicz\inst{\ref{inst4}}
  \and F.~Bensch\inst{\ref{inst5}}
  \and M.~I.~B\l{}\k{e}cka\inst{\ref{inst4}}
  \and M.~Emprechtinger\inst{\ref{inst3}}
  \and T.~Encrenaz\inst{\ref{inst2}}
  \and E.~Jehin\inst{\ref{inst6}}
  \and M.~K\"uppers\inst{\ref{inst7}}
  \and L.-M.~Lara\inst{\ref{inst8}}
  \and E.~Lellouch\inst{\ref{inst2}}
  \and B.~M.~Swinyard\inst{\ref{inst9}}
  \and B.~Vandenbussche\inst{\ref{inst10}}
  \and E.~A.~Bergin\inst{\ref{inst11}}
  \and G.~A.~Blake\inst{\ref{inst3}}
  \and J.~A.~D.~L.~Blommaert\inst{\ref{inst10}}
  \and J.~Cernicharo\inst{\ref{inst12}}
  \and L.~Decin\inst{\ref{inst10},\ref{inst23}}
  \and P.~Encrenaz\inst{\ref{inst13}}
  \and T.~de~Graauw\inst{\ref{inst18},\ref{inst24},\ref{inst14}}
  \and D.~Hutsem\'ekers\inst{\ref{inst6}}
  \and M.~Kidger\inst{\ref{inst15}}
  \and J.~Manfroid\inst{\ref{inst6}}
  \and A.~S.~Medvedev\inst{\ref{inst1}}
  \and D.~A.~Naylor\inst{\ref{inst16}}
  \and R.~Schieder\inst{\ref{inst17}}
  \and D.~Stam\inst{\ref{inst18}}
  \and N.~Thomas\inst{\ref{inst19}}
  \and C.~Waelkens\inst{\ref{inst10}}
  \and R.~Szczerba\inst{\ref{inst21}}
  \and P.~Saraceno\inst{\ref{inst22}}
  \and A.~M.~Di~Giorgio\inst{\ref{inst22}}
  \and S.~Philipp\inst{\ref{inst20}}
  \and T.~Klein\inst{\ref{inst20}}
  \and V.~Ossenkopf\inst{\ref{inst17},\ref{inst18}}
  \and P.~Zaal\inst{\ref{inst18}}
  \and R.~Shipman\inst{\ref{inst18}}
  }

\institute{
  Max-Planck-Institut f\"ur Sonnensystemforschung, 37191
  Katlenburg-Lindau, Germany\label{inst1}
  \and LESIA, Observatoire de Paris, 5 place Jules Janssen, 92195
    Meudon, France\label{inst2}
  \and California Institute of Technology, Pasadena, CA 91125, USA\label{inst3}
  \and Space Research Centre, Polish Academy of Sciences, Warsaw,
    Poland\label{inst4}
  \and DLR, German Aerospace Centre, Bonn-Oberkassel, Germany\label{inst5}
  \and Institute d'Astrophysique et de Geophysique, Universit\'e de Li\`ege,
    Belgium\label{inst6}
  \and Rosetta Science Operations Centre, European Space Astronomy
    Centre, European Space Agency, Spain\label{inst7}
  \and Instituto de Astrof\'isica de Andaluc\'ia (CSIC), Spain\label{inst8}
  \and STFC Rutherford Appleton Laboratory, Harwell Innovation
    Campus, Didcot, OX11 0QX, UK\label{inst9}
  \and Instituut voor Sterrenkunde, Katholieke Universiteit Leuven,
    Belgium\label{inst10}
  \and Astronomy Department, University of Michigan, USA\label{inst11}
  \and Laboratory of Molecular Astrophysics, CAB-CSIC, INTA, Spain\label{inst12}
  \and Sterrenkundig Instituut Anton Pannekoek, University of Amsterdam,
    Science Park 904, 1098 Amsterdam, The Netherlands\label{inst23}
  \and LERMA, Observatoire de Paris, France\label{inst13}
  \and SRON Netherlands Institute for Space Research, Landleven 12, 9747
    AD, Groningen, The Netherlands\label{inst18}
  \and Leiden Observatory, University of Leiden, The Netherlands\label{inst24}
  \and Joint ALMA Observatory, Chile\label{inst14}
  \and \herschel{} Science Centre, European Space Astronomy
    Centre, European Space Agency, Spain\label{inst15}
  \and Department of Physics and Astronomy, University of Lethbridge,
    Canada\label{inst16}
  \and KOSMA, I. Physik. Institut, Universit\"at zu K\"oln, Z\"ulpicher Str.
    77, D 50937 K\"oln, Germany\label{inst17}
  \and Physikalisches Institut, University of Bern, Switzerland\label{inst19}
  \and N. Copernicus Astronomical Center, Rabianska 8, 87-100, Torun,
    Poland\label{inst21}
  \and Istituto Fisica Spazio Interplanetario-INAF, Via Fosso del
    Cavaliere 100, 00133 Roma, Italy\label{inst22}
  \and Max-Planck-Institut f\"ur Radioastronomie, Auf dem H\"ugel 69, 53121
    Bonn, Germany\label{inst20}
  }

\date{Received May 31, 2010; accepted}

\abstract{The high spectral resolution and sensitivity of
\herschel{}/HIFI allows for the detection of multiple rotational water
lines and accurate determinations of water production rates in comets.
In this letter we present HIFI observations of the fundamental \transi{}
(557 GHz) ortho and \transii{} (1113 GHz) para rotational transitions of
water in comet \wild{} acquired in February 2010. We mapped the extent
of the water line emission with five point scans. Line profiles are
computed using excitation models which include excitation by collisions
with electrons and neutrals and solar infrared radiation.  We derive a
mean water production rate of $\mean \times 10^{28}$ molecules \s{} at a
heliocentric distance of 1.61 AU about 20 days before perihelion, in
agreement with production rates measured from the ground using
observations of the 18-cm OH lines.  Furthermore, we constrain the
electron density profile and gas kinetic temperature, and estimate the
coma expansion velocity by fitting the water line shapes.}

\keywords{Comets: individual: \wild{} --
      molecular processes --
      radiative transfer --
      submillimetre --
      techniques: spectroscopic
  }

\maketitle

\section{Introduction}

Water is the main volatile constituent of cometary comae.  Thus it is
crucial to determine the water production rate to understand cometary
activity and determine the relative abundances of other volatiles in the
coma \citep[see e.g.][]{1987A&A...181..169B}.  Water molecules in
cometary atmospheres are excited by collisions with neutrals and
electrons, and radiative pumping of the fundamental vibrational levels
by the solar infrared radiation.  The fundamental \transi{} ortho-water
transition at \ortho{} GHz is one of the strongest lines in cometary
spectra.  This transition has been detected using the Submillimeter Wave
Astronomical Satellite (SWAS) \citep{2000ApJ...539L.151N}, and
subsequently with Odin
\citep{2003A&A...402L..55L,2007P&SS...55.1058B,2009A&A...501..359B} and
the \herschel{} Space Observatory \citep{2010Hartogh}.  \herschel{}
provided the first detection of the $2_{12}$--$1_{01}$ (1669.905 GHz)
ortho and $1_{11}$--$0_{00}$ (\para{} GHz) para transitions of water in
cometary atmospheres \citep{2010Hartogh}.  The continuum emission in
comet C/2006 W3 (Christensen) has been measured by \herschel{}
\citep{2010arXiv1005.1592B}.

Comet \wild{} is a Jupiter family comet that was the target of NASA's
\stardust{} sample return mission \citep{2004Sci...304.1764B}.
Thousands of particles in the submillimeter size range were collected
for laboratory study during the \stardust{} flyby on January 2, 2004
\citep{2006Sci...314.1711B}.  Optical images of the nucleus
obtained by the navigation camera show very complex
topographic features \citep{2004Sci...304.1764B}.  Gas production rates
for several molecules have been monitored at multiple apparitions
\citep{1992Icar...98..115O,2001Icar..152..268M,1999Icar..141..331F,2005Icar..173..533F}.
A maximum water production rate of $Q_\water{} \sim 1.1 \times 10^{28}\
\s$ is estimated from narrowband photometry of OH emission in the
near-UV \citep{2005Icar..173..533F}.

\begin{DIFnomarkup} 
The Heterodyne Instrument for the Far-Infrared (HIFI) onboard
\herschel{} is well suited to observe water vapor in cometary coma with
high spectral resolution \citep{2010HIFI,2010arXiv1005.5331P}.  Comet \wild{}
was observed with HIFI on February 1--4, 2010, in the framework of the
\herschel{} guaranteed time key project
``\href{http://www.mps.mpg.de/projects/herschel/HssO/index.htm}{Water
and related chemistry in the Solar System}'' \citep{2009P&SS...57.1596H}.
\end{DIFnomarkup} 

\section{HIFI observations}
\label{sec:observations}

Comet 81P/Wild 2 passed its perihelion on February 22.7 UT at
$\rh=1.5981$ AU from the Sun. The \herschel{} observations took place in
early February about 20 days before perihelion, at heliocentric distance
$\rh \simeq 1.6$ AU and a distance of $\Delta \simeq 0.9$ AU from
\herschel{}.  The comet could not be observed after February 14 due to
visibility constraints.  HIFI can observe two polarizations
simultaneously. With a spectral resolution of 1.1 MHz (Wide Band
Spectrometer; WBS) or 140 KHz (High Resolution Spectrometer; HRS) HIFI
can resolve spectrally cometary line shapes and asymmetries
\citep{2010Hartogh}.

\begin{DIFnomarkup}
Table~\ref{table:1} shows a summary of the HIFI observations of the
\transi{} (557 GHz, HIFI bands 1a and 1b) and \transii{} (1113 GHz, HIFI
band 4b)  water lines in comet \wild{}. The position of the comet and
its relative motion with respect to \herschel{} were calculated using
the \href{http://ssd.jpl.nasa.gov/?horizons}{JPL's HORIZONS system}.
\end{DIFnomarkup}
The \herschel{} telescope has a diameter of 3.5 m, with the
corresponding HIFI Half Power Beam Widths (HPBW) of $19.2\arcsec$ and
$38.1\arcsec$ at \paras{}, and \orthos{} GHz, respectively. The
corresponding beam sizes projected on the comet are $1.5$ and $2.2
\times10^{4}$ km, respectively.
Observations were conducted in the standard dual beam switch (DBS) cross
map observing mode with a chopper speed of 4 Hz and a separation of the
cross positions corresponding to a Nyquist sampling of the beam size
(22$\arcsec$, 19$\arcsec$, and 10$\arcsec$ for observations in bands 1a,
1b, and 4b, respectively). One axis of the cross was aligned along the
Sun direction at PA = $111\degr$. The reference OFF position was
$3\arcmin$ from the position of the comet.  This observing mode was not
released at the time of the observations.

\begin{table*}
  \caption{HIFI observations of water vapour in comet \wild{} and
  derived water production rates. The observations were acquired
  in February 2010 using the standard fast DBS cross map observing mode.
  The intensity, velocity shift and $Q_\water{}$ correspond to the
  average of the central points in the cross maps.
  }
  \label{table:1} \centering
  \begin{tabular}{c c c c c c c c r@{}l c}
    \hline\hline Obs.\ ID & UT start date & $\Delta$ & $r_\mathrm{h}$ & Band &
    Frequency & Integration time & Intensity &
    \multicolumn{2}{c}{Velocity shift} & $Q_\water{}$\tablefootmark{a} \\
    & [mm/dd.ddd] & [AU] & [AU] & & [GHz] & [s] & [K \kms] &
    \multicolumn{2}{c}{[\ms]} & [$10^{28}\ \mathrm{s}^{-1}$] \\
    \hline
    1342190098 & 02/1.443 & 0.949 & 1.612 & 1a & 556.9360 & 684 & $1.807 \pm 0.012$ & $4\,\pm\,$ & 5 & $1.00 \pm 0.05$\\
1342190185 & 02/4.235 & 0.927 & 1.609 & 1b & 556.9360 & 748 & $2.077 \pm 0.015$ & $-10\,\pm\,$ & 5 & $1.13 \pm 0.06$\\
1342190231 & 02/2.134 & 0.943 & 1.611 & 4b & 1113.3430 & 390 & $1.997 \pm 0.053$ & $-58\,\pm\,$ & 21 & $1.07 \pm 0.06$\\
1342190232 & 02/2.151 & 0.943 & 1.611 & 4b & 1113.3430 & 366 & $1.594 \pm 0.052$ & $-41\,\pm\,$ & 26 & $0.86 \pm 0.05$\\

    \hline
  \end{tabular}
  \tablefoottext{a}{ Water production rates computed with $T$ = 40 K and
  $\xne = 0.2$. The error bar includes a 5\% calibration uncertainty in
  the sideband gain ratio.}
 
\end{table*}

\section{Data analysis}
\label{sec:analysis}

\subsection{Water line emission}

The data were reduced to the level 2 products using the standard
{\it Herschel} Interactive Processing Environment (HIPE) 3.0
pipeline for HIFI \citep{2010HIFI}.  Integrated line
intensities and velocity offsets in the comet rest frame are given
in Table~\ref{table:1} for the central point in the maps.
Line intensities are calculated from the weighted averages of
spectra in horizontal (H) and vertical (V) polarizations. We scaled the
main beam brightness temperature using the beam efficiency of the
\herschel{} telescope of 0.7 and 0.75 for the \paras{} and \orthos{} GHz
lines respectively \citep[see e.g.][]{2010Hartogh}.  The uncertainties
in line intensity, velocity shift and water production rates are 1-$\sigma$ 
statistical uncertainties. Uncertainties in the sideband gain ratio,
which reach 5\% \citep{2010Roelfsema}, are considered in the data
analysis.

Figure~\ref{fig:557} shows the HRS spectrum of the \transi{} ortho-water
line at \orthos{} GHz observed on February 4.24 UT at the center of the
map. The central spectrum of the map of the \transii{} para-water line
at \paras{} GHz, averaging the two observations to increase the
signal-to-noise ratio, is shown in Fig.~\ref{fig:1113}. These spectra
are uniformly weighted averages of the H and V polarizations.  They
correspond also to the average of the central points of the horizontal
and vertical drifts of the cross map.  The frequency scale of the
observed spectra was corrected for the geocentric velocity of the comet
and the spacecraft orbital velocity. The lines are optically thick and
slightly asymmetric due to self-absorption effects in the foreground
coma. We estimate a mean expansion velocity $v_\mathrm{exp} = 0.6\ \kms$
from the width of the \orthos{} GHz line \citep{2007P&SS...55.1058B}.
Expansion velocities in the range $v_\mathrm{exp} = 0.5-0.8\ \kms$ are
typical for weak comets.

\pgfplotsset{every tick/.append style={color=black}}
\pgfplotsset{every tick/.append style={thin}}

\begin{figure}
  \centering
  \pgfplotsset{ortho style/.style=
    {xmin=-5, xmax=5, ymin=-0.2, ymax=2.1, width=\hsize, height=7cm, 
      minor x tick num=3, minor y tick num=4}}
  \pgfplotsset{every axis y label/.append style={yshift=-0pt}}
  \begin{tikzpicture}
    \begin{axis}[
      ylabel={$T_\mathrm{mB}$ [K]}, xlabel={$v$ [km s$^{-1}$]},
      ortho style]
      \addplot[mark=none,const plot mark right]
      table[x index=0,y index=1]
       {figures/1342190185_HSO_HRS_center.txt};
      \addplot[mark=none,blue,dashed]
      table[x index=0,y index=1]
      {figures/185.txt};
    \end{axis}
  \end{tikzpicture}
  \caption{Central point in the DBS map of the ortho-water
  line at \ortho{} GHz in comet \wild{} obtained by the HRS on February
  4.24 UT.
  The velocity scale is given with respect to the comet rest frame
  with a resolution of $\sim 32\ \ms$.
  A synthetic line profile for isotropic outgassing with
  $v_\mathrm{exp} = 0.6\ \kms$, $T = 40\ \mathrm{K}$, and $\xne{} = 0.2$
  is shown by the dashed line.
  }
  \label{fig:557}
\end{figure}
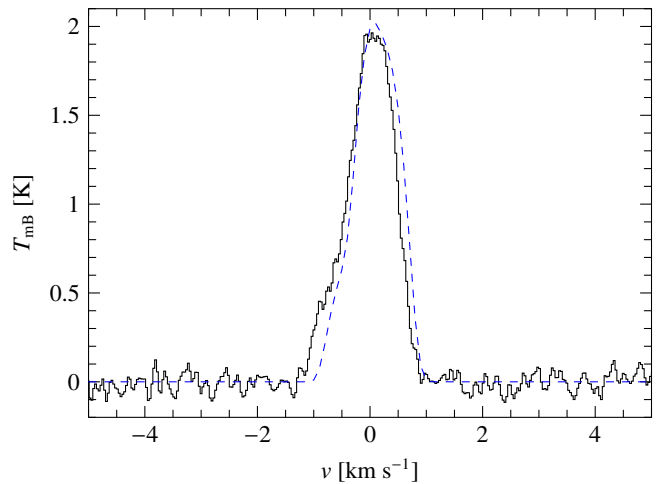

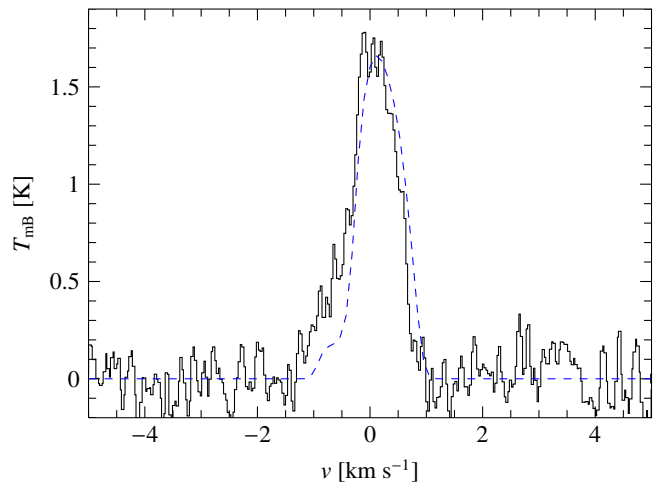
\begin{figure}
  \centering
  \pgfplotsset{para style/.style=
    {xmin=-5, xmax=5, ymin=-0.2, ymax=1.9, height=7cm, width=\hsize,
    minor x tick num=3, minor y tick num=4}}
  \begin{tikzpicture}
    \begin{axis}[xlabel={$v$ [km s$^{-1}$]},
      ylabel={$T_\mathrm{mB}$ [K]}, para style, name=p0]
      \addplot[mark=none,const plot mark right]
      table[x index=0,y index=1]
      {figures/fig2_smooth_HSO_HRS.txt};
      \addplot[mark=none,blue,dashed]
      table[x index=0,y index=1]
      {figures/232.txt};
    \end{axis}
  \end{tikzpicture}
  \caption{
  Average of the HRS observations of the para-water line at \para{} GHz
  towards the nucleus of comet \wild{} obtained on February 2.13 UT
  and 2.15 UT.
  The velocity resolution is $\sim 32\ \ms$ after smoothing.  
  The dashed line shows the best fit profile 
  for the same parameters as in Fig.~\ref{fig:557}.
  }
  \label{fig:1113}
\end{figure}

\begin{figure}
  \centering
  \pgfplotsset{map style/.style=
    {xmin=-5, xmax=5, ymin=-0.2, ymax=2., height=.47\hsize,
    width=.47\hsize, minor x tick num=3, minor y tick num=4}}
  \begin{tikzpicture}
    \begin{axis}[map style, name=p18,
    yticklabel=\empty,xticklabel=\empty]
      \addplot[mark=none,const plot mark right]
      table[x index=0,y index=1]
      {figures/1342190098_HSO_HRS_center.txt};
    \end{axis}
    \begin{axis}[map style, at={(p18.west)}, anchor=east,
      ylabel={$T_\mathrm{mB}$ [K]}, xlabel={$v$ [\kms]}, 
      title =$\sun \leftarrow$]
      \addplot[mark=none,const plot mark right]
      table[x index=0,y index=1]
      {figures/1342190098_HSO_HRS_10.txt};
    \end{axis}
    \begin{axis}[map style, at={(p18.east)}, anchor=west,
      xlabel={$v$ [\kms]}, 
      yticklabel=\empty]
      \addplot[mark=none,const plot mark right]
      table[x index=0,y index=1]
      {figures/1342190098_HSO_HRS_0.txt};
    \end{axis}
    \begin{axis}[map style, at={(p18.south)}, anchor=north,
      ylabel={$T_\mathrm{mB}$ [K]}, xlabel={$v$ [\kms]}, 
      yticklabels={0,0,0.5,1,1.5}]
      \addplot[mark=none,const plot mark right]
      table[x index=0,y index=1]
      {figures/1342190098_HSO_HRS_22.txt};
    \end{axis}
    \begin{axis}[map style, at={(p18.north)}, anchor=south,
      ylabel={$T_\mathrm{mB}$ [K]},
      xticklabel=\empty]
      \addplot[mark=none,const plot mark right]
      table[x index=0,y index=1]
      {figures/1342190098_HSO_HRS_14.txt};
    \end{axis}
  \end{tikzpicture}
  \caption{
  DBS cross-map of the ortho-water \transi{} transition 
  obtained with the HRS on February 1.44 UT.
  Offset positions are $\sim 22\arcsec$ apart. 
  The velocity resolution is $\sim 32\ \ms$.  
  The Sun direction is indicated.
  }
  \label{fig:crossmap}
\end{figure}
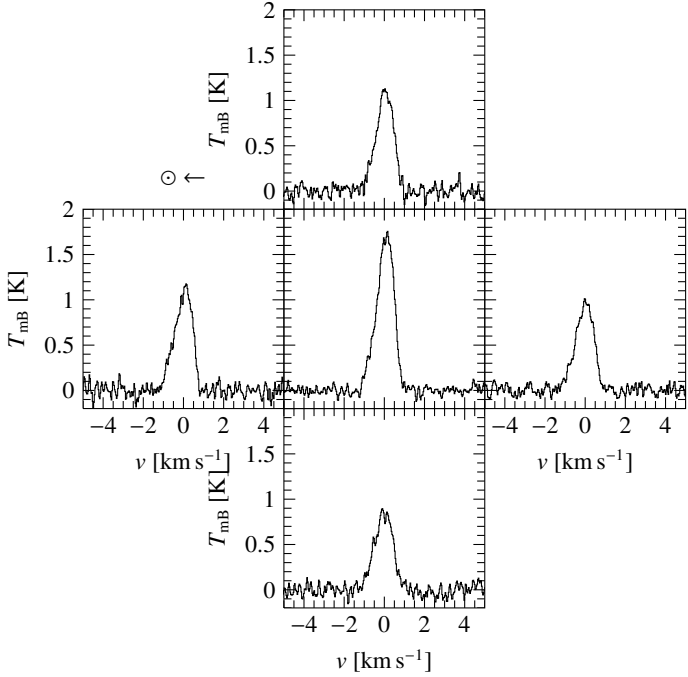

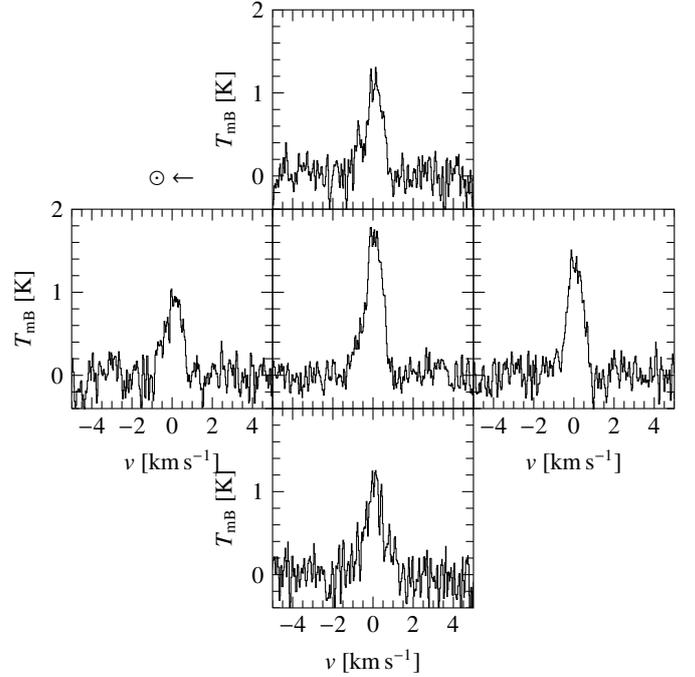
\begin{figure}
  \centering
  \pgfplotsset{map style/.style=
    {xmin=-5, xmax=5, ymin=-0.4, ymax=2., height=.47\hsize, width=.47\hsize,
      minor x tick num=3, minor y tick num=4}}
  \begin{tikzpicture}
    \begin{axis}[map style, name=p12,
      yticklabel=\empty, xticklabel=\empty]
      \addplot[mark=none,const plot mark right]
      table[x index=0,y index=1]
      {figures/fig2_smooth_HSO_HRS.txt};
    \end{axis}
    \begin{axis}[map style, at={(p12.west)}, anchor=east,
      ylabel={$T_\mathrm{mB}$ [K]}, xlabel={$v$ [\kms]}, 
      title =$\sun \leftarrow$]
      \addplot[mark=none,const plot mark right]
      table[x index=0,y index=1]
      {figures/1113_HSO_HRS_smooth_2.txt};
    \end{axis}
    \begin{axis}[map style, at={(p12.east)}, anchor=west,
      xlabel={$v$ [\kms]}, 
      yticklabel=\empty]
      \addplot[mark=none,const plot mark right]
      table[x index=0,y index=1]
      {figures/1113_HSO_HRS_smooth_0.txt};
    \end{axis}
    \begin{axis}[map style, at={(p12.south)}, anchor=north,
      ylabel={$T_\mathrm{mB}$ [K]}, xlabel={$v$ [\kms]}, 
      yticklabels={0,0,1}]
      \addplot[mark=none,const plot mark right]
      table[x index=0,y index=1]
      {figures/1113_HSO_HRS_smooth_5.txt};
    \end{axis}
    \begin{axis}[map style, at={(p12.north)}, anchor=south,
      ylabel={$T_\mathrm{mB}$ [K]},
      xticklabel=\empty]
      \addplot[mark=none,const plot mark right]
      table[x index=0,y index=1]
      {figures/1113_HSO_HRS_smooth_3.txt};
    \end{axis}
  \end{tikzpicture}
  \caption{
  Average of the cross-maps of the \transii{} para-water
  line obtained with the HRS on February 2.13 and 2.15 UT.
  The offset spacing is $\sim 10\arcsec$ and
  the effective velocity resolution after smoothing is $\sim 32\ \ms$.  
  The Sun direction is indicated.
  }
  \label{fig:1113map}
\end{figure}

Mapping observations at \orthos{} and \paras{} GHz are shown in
Figs.~\ref{fig:crossmap} and \ref{fig:1113map},
and {\it online} Fig.~\ref{fig:185map}.

\onlfig{5}{
\begin{figure}
  \centering
  \pgfplotsset{map style/.style=
    {xmin=-5, xmax=5, ymin=-0.2, ymax=2., height=.47\hsize, width=.47\hsize,
      minor x tick num=3, minor y tick num=4}}
  \begin{tikzpicture}
    \begin{axis}[map style, name=p12,
      yticklabel=\empty, xticklabel=\empty]
      \addplot[mark=none,const plot mark right]
      table[x index=0,y index=1]
      {figures/1342190185_HSO_HRS_center.txt};
    \end{axis}
    \begin{axis}[map style, at={(p12.west)}, anchor=east,
      xlabel=\vlabel,ylabel=\tlabel, title =$\sun \leftarrow$]
      \addplot[mark=none,const plot mark right]
      table[x index=0,y index=1]
      {figures/1342190185_HSO_HRS_12.txt};
    \end{axis}
    \begin{axis}[map style, at={(p12.east)}, anchor=west,
      xlabel=\vlabel, yticklabel=\empty]
      \addplot[mark=none,const plot mark right]
      table[x index=0,y index=1]
      {figures/1342190185_HSO_HRS_0.txt};
    \end{axis}
    \begin{axis}[map style, at={(p12.south)}, anchor=north,
      xlabel=\vlabel,ylabel=\tlabel,yticklabels={0,0,0.5,1,1.5}]
      \addplot[mark=none,const plot mark right]
      table[x index=0,y index=1]
      {figures/1342190185_HSO_HRS_27.txt};
    \end{axis}
    \begin{axis}[map style, at={(p12.north)}, anchor=south,
      ylabel=\tlabel,xticklabel=\empty]
      \addplot[mark=none,const plot mark right]
      table[x index=0,y index=1]
      {figures/1342190185_HSO_HRS_17.txt};
    \end{axis}
  \end{tikzpicture}
  \caption{
  DBS cross-map of the \transi{} ortho-water line obtained
  with the HRS on February 4.24 UT.  The offset spacing is $\sim
  19\arcsec$ and the velocity resolution is $\sim 32\ \ms$.  
  The Sun direction is indicated.
  }
  \label{fig:185map}
\end{figure}
}

\subsection{Radiative transfer modelling}

Molecular excitation in the outer coma is dominated by
collisions with electrons and infrared fluorescence by solar
radiation. We use a spherically symmetric Monte Carlo radiative
transfer numerical code to compute the populations of the water
rotational levels as function of the distance from the nucleus
\citep[see][and references therein]{2010Hartogh}. A constant gas
temperature of $T
 = 40$ K is assumed. The ortho-to-para water ratio in comets varies
from 2.5 to 3 \citep[e.g.][and references
therein]{2007ApJ...661L..97B}. We assume a value of 3. The scaling
factor of the electron density profile with respect to a nominal
profile deduced from 1P/Halley {\it in situ} data is chosen to be
$\xne = 0.2$ and 1. The former value was found to best explain the
brightness distribution of the \transi{} line obtained from
mapping observations \citep{2007P&SS...55.1058B,2010Hartogh}.

The radiative transfer equation is solved along different lines of sight
through the coma covering $2.5\times\mathrm{HPBW}$, and the beam
averaged emission spectra are computed.  We obtain the water production
rate by comparing the calculated and observed line intensities. The
synthetic line profiles agree approximately with the observed line
shapes (Figs.~\ref{fig:557}--\ref{fig:1113}).  The differences (overall
shift and excess emission in the blue wing of the profiles) suggest
preferential outgassing towards the Sun and a day/night asymmetry in the
gas velocity field.

\subsection{Water production rates}

\begin{figure}
  \centering
  \includegraphics[angle=270,width=0.5\textwidth]{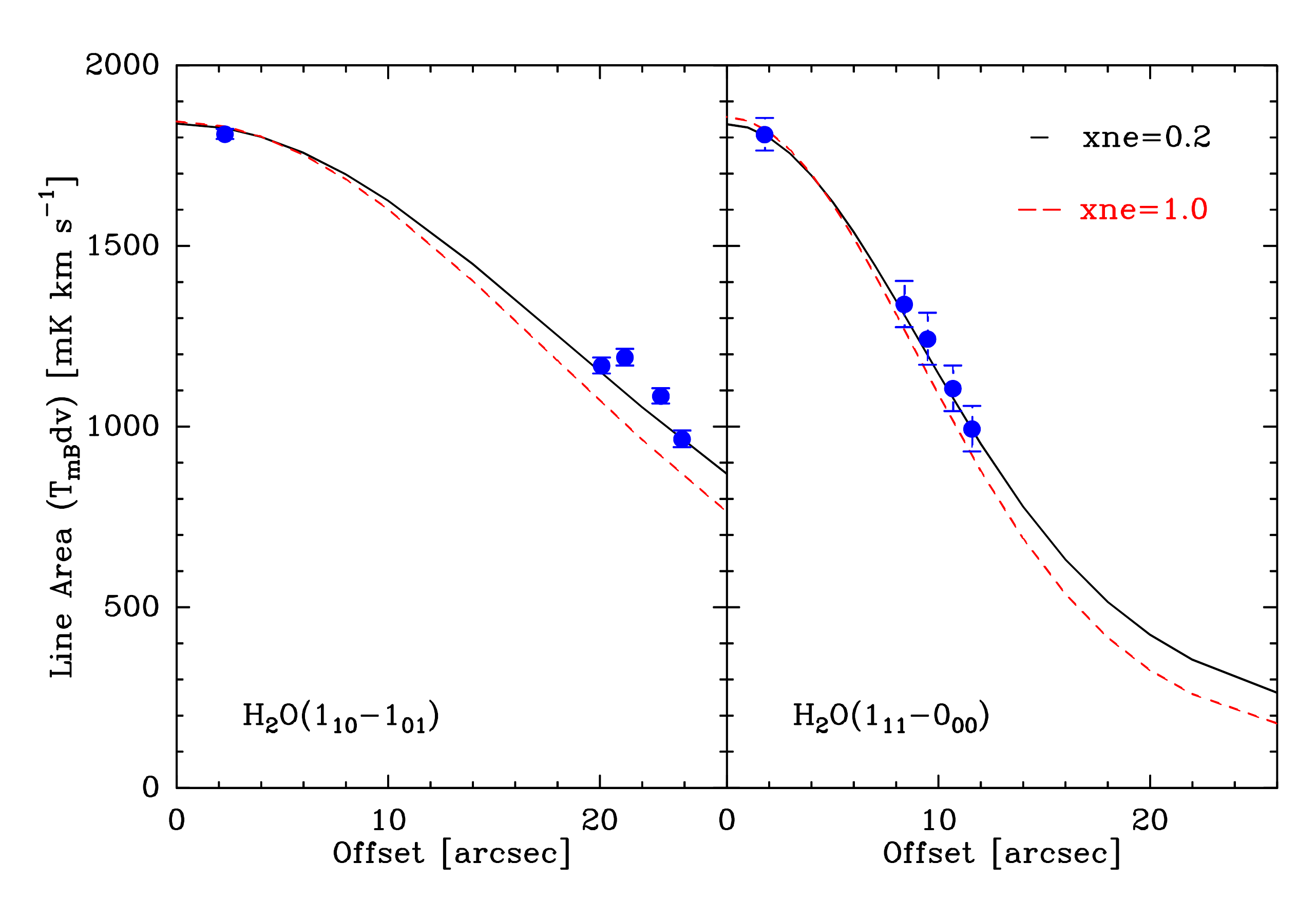}
  \caption{Intensities of the \transi{} (Feb. 1.44 UT, left panel) and \transii{}
  (Feb. 2.13-2.15 UT, right panel) lines as a function of position offset from the
  estimated position of the peak brightness.  Model calculations for
  $\xne = 0.2$ and 1 are shown with a black solid line and
  red dashed line, respectively. We assumed $T$ = 40 K and
  $\vexp{} = 0.6\ \kms$.}
  \label{fig:fit-map}
\end{figure}

Mapping observations can be used to constrain excitation
parameters such as the neutral gas kinetic temperature and
electron density scaling factor. The spacing of $10\arcsec$
and $22\arcsec$ between the observed positions corresponds to
distances of $\sim 1.2\times10^4$ and $2.4\times10^4\ \mathrm{km}$
at the distance of the comet, respectively, where the excitation of the water
rotational levels is controlled both by collisions with electrons
and infrared fluorescence \citep[see Fig. 5 in][]{2010Hartogh}.
Figure~\ref{fig:fit-map} shows the evolution of the line area of
the 557 and 1113 GHz lines as a function of the position
offset. As shown in Fig.~\ref{fig:fit-map}, model calculations
performed with $T$ = 40 K and $\xne = 0.2$ provide a satisfactory
fit to the observed radial brightness profiles. For a larger
electron content ($\xne = 1$), the line intensities at the offset
positions are lower than observed. Production rates for $\xne{}
= 0.2$ are consistent with values obtained from OH observations
(Crovisier et al.\ in preparation), while those for
$\xne{} = 1$ are about a factor of two lower.

The water production rates derived from the 557 and 1113 GHz lines
are consistent with each other and in the range $\range
\times10^{28}\ \mathrm{s}^{-1}$. Model calculations show that the
$I$(557 GHz)/$I$(1113 GHz) intensity ratio does not depend
strongly on the $\xne$ parameter. Gas kinetic temperatures
$T$ in the range 30--50 K provide a satisfactory fit of the
measured intensity ratio, averaging the lines observed at the
different dates and considering uncertainties in flux calibration.
Lower temperatures (by $\sim$ 15 K) are retrieved when the
ortho-to-para ratio is taken equal to 2.5. Such low $T$ values are
typical for comets of relatively low activity. However, these
temperature retrievals are possibly affected by the
non-simultaneity of the observations.

\subsection{Comparison with observations of water photodissociation products}

In support of the \herschel{} observations, the OH lines at 18-cm were
observed with the Nan\c{c}ay radio telescope from November 16, 2009 to
February 14, 2010 (Crovisier et al.\ in preparation). The methods of
observation and analysis are described in \citet{2002A&A...393.1053C}.
The comet could be detected only after integrating over many days in
December and January.  It could not be detected close to perihelion
because the OH maser inversion was small at that time.  Water is the
only significant parent molecule of the OH radical in cometary
atmospheres and the lifetimes of the two species are well known,
allowing for a determination of the water production rate \citep[see
e.g.][]{1993Icar..105..235C}.  The retrieved water production rates,
including the HIFI retrievals, are shown in Fig.~\ref{fig:qh2o}.

Figure~\ref{fig:qh2o} includes water production rates measured for the
1997 apparitions from OH narrowband photometry
\citep{2005Icar..173..533F},  
spectroscopy of the $^1$D line of atomic oxygen \citep{1999Icar..141..331F} and
Lyman-$\alpha$ measurements made with the SWAN instrument onboard SOHO
\citep{2001Icar..152..268M}. The latter water production rates are about a
factor of two higher than those obtained from narrowband photometry of
OH emission \citep{2005Icar..173..533F}.
The higher production rates measured
80--20 days before perihelion may indicate a seasonal effect, with a
water production rate peak at $\sim 40$--60 days pre-perihelion
\citep[see e.g.][]{2008A&A...490..393S}.

\begin{figure}
  \begin{tikzpicture}
  \pgfplotsset{qh2o style/.style=
    {width=\hsize, xmin=-120,xmax=80}}
  \begin{axis}[xlabel={Days from perihelion},
    ylabel={$Q_\water{}$ [$10^{28}\ \s$]}, 
    qh2o style,
    minor x tick num=1, minor y tick num=4,
    ytick scale label code/.code={},
    ymin=0, ymax=2.5e28]
    \addplot+[only marks]
      plot[error bars/.cd,y dir=both,y explicit, x dir=both,x explicit]
      table[x index=0,x error index=1,y index=2,y error index=3]
      {figures/nancay.txt};
    \addplot+[only marks]
      plot[error bars/.cd,y dir=both,y explicit]
      table[x index=0,y index=1,y error index=2]
      {figures/qh2o.txt};
    \addplot+[only marks,mark=triangle]
      plot[error bars/.cd,y dir=both,y explicit relative]
      table[x index=0,y index=1,y error index=2]
      {figures/farnham1997.txt};
    \addplot+[only marks,mark=diamond,color=green]
      plot[error bars/.cd,y dir=both,y explicit]
      table[x index=0,y index=1,y error index=2]
      {figures/fink1997.txt};
    \addplot+[only marks,mark=square,color=magenta]
      plot[error bars/.cd,y dir=both,y explicit]
      table[x index=0,y index=1,y error index=2]
      {figures/makin1997.txt};
    \addplot+[only marks,mark=o,color=cyan,solid]
      plot[error bars/.cd,y dir=both,y explicit,x dir=both,x explicit]
      table[x index=0,x error index=1,y index=2,y error index=3]
      {figures/nancay1997.txt};
  \end{axis}
  \end{tikzpicture}
  \caption{ Water production rates with 1--$\sigma$ uncertainties in
  comet \wild{} as a function of time from perihelion. HIFI measurements
  are shown by squares. Filled circles represent OH 18-cm line observations
  with the Nan\c{c}ay radio telescope.  Water production rates for the
  1997 apparition from \citet{2002A&A...393.1053C},
  \citet{2005Icar..173..533F}, \citet{1999Icar..141..331F} and
  \citet{2001Icar..152..268M} are shown by empty circles, triangles,
  diamonds and squares, respectively.
  }
  \label{fig:qh2o}
\end{figure}

\section{Conclusions}
\label{sec:conclusions}

Although comet \wild{} has been extensively observed from the ground, it
is still poorly observed at sub-millimeter wavelengths.  \wild{} was
observed with HIFI in the period 1-4 February 2010 at 1.6 AU
heliocentric distance and 0.95 AU from \herschel{}. The fundamental
ortho- and para-water rotational transitions at \ortho{} GHz and \para{}
GHz were detected at high spectral resolution. DBS cross-maps were
obtained to study the excitation conditions throughout the coma.  We
calculate a water production rate in the range $\range \times10^{28}\
\s$ using a radiative transfer code which includes collisional effects
and infrared fluorescence by solar radiation. Water production rates are
in good agreement with those derived from ground-based observations of
the OH 18-cm emission at the Nan\c{c}ay radio telescope about 30 days
before perihelion.

\begin{acknowledgements}
HIFI has been designed and built by a consortium of institutes and university
departments from across Europe, Canada and the United States under the
leadership of SRON Netherlands Institute for Space Research, Groningen, The
Netherlands and with major contributions from Germany, France and the US.
Consortium members are: Canada: CSA, U.Waterloo; France: CESR, LAB, LERMA,
IRAM; Germany: KOSMA, MPIfR, MPS; Ireland, NUI Maynooth; Italy: ASI, IFSI-INAF,
Osservatorio Astrofisico di Arcetri-INAF; Netherlands: SRON, TUD; Poland: CAMK,
CBK; Spain: Observatorio Astron\'omico Nacional (IGN), Centro de Astrobiolog\'ia
(CSIC-INTA). Sweden: Chalmers University of Technology - MC2, RSS \& GARD,
Onsala Space Observatory, Swedish National Space Board, Stockholm University -
Stockholm Observatory; Switzerland: ETH Z\"urich, FHNW; USA: Caltech, JPL, NHSC.
HIPE is a joint development by the \herschel{} Science Ground Segment Consortium,
consisting of ESA, the NASA \herschel{} Science Center, and the HIFI, PACS and
SPIRE consortia.
This development has been supported by national funding agencies: CEA, CNES,
CNRS (France); ASI (Italy); DLR (Germany).
Additional funding support for some instrument activities has been provided by
ESA.
Support for this work was provided by NASA through an award issued by
JPL/Caltech.
DCL is supported by the NSF, award AST-0540882 to the Caltech
Submillimeter Observatory.
SS, MB and MIB are supported by the Polish Ministry of Education and Science
(MNiSW).
\end{acknowledgements}

\bibliographystyle{aa}
\bibliography{ads,preprints}

\end{document}